\documentstyle[eqsecnum,aps]{revtex}

\newcommand{\hn}{\hat \nabla}

\newcommand{\gab}{g_{ab}}

\newcommand{\gnm}{g^{nm}}
\newcommand{\hgnm}{\hat g^{nm}}
\newcommand{\Tab}{T_{ab}}

\begin{document}

\draft

\title{ Conformal classes of Brans-Dicke gravity }
 
\author{ Israel Quiros\thanks{israel@mfc.esivc.colombus.cu}}
\address{ Departamento de Fisica. Universidad Central de Las Villas. Santa Clara. CP: 54830 Villa Clara. Cuba }

\date{\today}

\maketitle

\begin{abstract}
A classification of Brans-Dicke theories of gravitation, based on the behaviour of the dimensionless gravitational coupling constant, is given. It is noted that the discussion takes place in the current literature, about which of the two distinguished conformal frames in which scalar-tensor theories of gravity can be formulated: the Jordan frame and the Einstein frame, is the physical one, may, in most cases, be meaningless for both frames may belong to the same conformal class. It is also noted that the Jordan frame formulation of Brans-Dicke gravity with ordinary matter nonminimally coupled, that is shown to be just the Jordan frame formulation of general relativity, is scale-invariant, unlike the situation with the Jordan frame representation of Brans-Dicke gravity with matter minimally coupled (the original formulation of Brans-Dicke theory), where the presence of nonzero mass ordinary matter breaks the scale-invariance of the theory.
\end{abstract}


Among all conformal mappings relating different scalar-tensor gravity (STG) theories, two frames are distinguished \cite{sok}: the Jordan frame (JF) and the Einstein frame (EF).

Some arguments have been raised against the physical equivalence of both formulations of STG theories (see Ref.\cite{sok,mag,fgn} and references therein), so it is of prime interest to know which frame is the physical one. This issue is being intensively discussed in the literature (see the review \cite{fgn}).

The most undesirable feature of EF STG theories is that, in this frame, the ordinary matter is nonminimally coupled to the scalar field so, in particular, test particles don't follow the geodesics of the geometry. For his part, the strongest objection against JF STG theories is that this formulation of STG leads to a negative definite, or indefinite kinetic energy for the scalar field, this implies that the theory does not have a stable ground state \cite{mae,gp}.

However, as it has been carefully noted in \cite{ss}, the terms with the second covariant derivatives of the scalar field on the right hand side of the JF field equation, contain the connection, and hence a part of the dynamical description of gravity. Santiago and Silbergleit \cite{ss} introduced a new connection that removes the gravitational dynamical terms from the right of the Einstein field equation, leading to the correct energy-momentum tensor for the scalar field. The scalar field energy density can now be made nonnegative by imposing some constraints to the coupling function and to the potential.

Yet another source of uncertainty respecting this subject was introduced by Magnano and Sokolowski \cite{sok,mag} when they studied the possibility of adding matter nonminimally coupled in the JF, such as to have the matter minimally coupled in the EF.

Our letter is aimed precisely at the study of this possibility for Brans-Dicke (BD) gravity (the prototype of STG theory) and, in connection with it, the classification of BD theories attending the different kind of behaviour of the dimensionless gravitational coupling constant $Gm^2 (\hbar = c = 1)$ that can arise. 

We shall begin with the JF formulation of BD gravity due to Brans and Dicke \cite{bdk}, that is based upon the Lagrangian density:

$$
L^I[g,\phi]=\frac{\sqrt{-g}}{16\pi}(\phi R - \frac{\omega}{\phi} \gnm \nabla_n \phi \nabla_m \phi) + L_{matter}[g]
\eqno{(1)}
$$

where $R$ is the curvature scalar, $\phi$ is the scalar BD field, $\omega$ is the BD coupling constant, and $L_{matter}[g]$ is the Lagrangian density of the ordinary matter minimally coupled to the scalar BD field.

The Lagrangian density conformally dual to (1) gives rise to the EF formulation of BD gravity \cite{dk}:

$$
L^I[\hat g,\hat \phi]=\frac{\sqrt{-\hat g}}{16\pi}(\hat R - (\omega + \frac{3}{2}) \hgnm \hn_n \hat \phi \hn_m \hat \phi) + \hat L_{matter}[\hat g,\hat \phi ]
\eqno{(2)}
$$

where $\hat R$ is the curvature scalar in the EF metric $\bf \hat g$, conformally dual to $\bf g$:

$$
\hat \gab = \phi \gab
\eqno{(3)}
$$

The scalar function $\hat \phi \equiv \ln \phi$ is the EF BD scalar field, and $\hat L_{matter}[\hat g,\hat \phi]$ is the Lagrangian density for the ordinary matter nonminimally coupled to the scalar field.

Magnano and Sokolowski \cite{sok,mag} studied the possibility of changing the coupling in (1); $L_{matter}[g] \rightarrow L_{matter}[g,\phi]$, while keeping intact the gravitational part:

$$
L^{II}[g,\phi]=\frac{\sqrt{-g}}{16\pi}(\phi R - \frac{\omega}{\phi} \gnm \nabla_n \phi \nabla_m \phi) + L_{matter}[g,\phi]
\eqno{(4)}
$$

In this case, unlike the original JF BD theory as formulated by Brans and Dicke \cite{bdk}, the ordinary matter is nonminimally coupled to the scalar field in the JF.

The EF of the theory based on (4) is given by the Lagrangian density conformally dual to it:

$$
L^{II}[\hat g,\hat \phi]=\frac{\sqrt{-\hat g}}{16\pi}(\hat R - (\omega + \frac{3}{2}) \hgnm \hn_n \hat \phi \hn_m \hat \phi) + \hat L_{matter}[\hat g]
\eqno{(5)}
$$

In this frame, test particles follow the geodesics of the geometry, so the inertial mass $m$ of a given material particle is constant over the spacetime manifold $M$. At the same time, the scalar field $\hat \phi$ is minimally coupled to the curvature (it is a distinctive feature of the EF), so the dimensional gravitational constant $G$ is a real constant, this means that the dimensionless gravitational coupling constant $Gm^2 (\hbar = c = 1)$ is constant on $M$ as well, i.e.; the strong equivalence principle (SEP) holds in the EF of the theory. Besides, since $Gm^2$ is a dimensionless constant, it is not affected by the conformal rescaling of the metric (eq.(3))\footnotemark \footnotetext{In fact it is a conformal transformation of the units of measurement \cite{dk}.}, so it is a constant in the JF of the theory as well.

This way, the distinctive feature of the theory given by $L^{II}[g,\phi]$ (and its conformally dual given by $L^{II}[\hat g,\hat \phi]$) is that the dimensionless gravitational coupling constant $Gm^2$ is a real constant over $M$. The class given by the pair \{$L^{II}[g,\phi], L^{II}[\hat g,\hat \phi]$\} we shall call {\it Class II BD} gravity, while the class given by the pair \{$L^I[g,\phi], L^I[\hat g,\hat \phi]$\} we shall call {\it Class I BD} gravity. The distinctive feature of this last class is that the dimensionless gravitational coupling constant changes from point to point in spacetime: $Gm^2 \sim \phi^{-1}$ \cite{bdk}. Such as in this case both the weak equivalence principle (WEP) and SEP hold, we can realize that the {\it Class II BD} theory is just general relativity (GR), so (4) leads to the JF GR.

The different formulations inside of a class are dual to each other in the sense that experiment can not differentiate between them. Actually, both the JF and the EF of the theory are connected by the conformal transformation of units (3) \cite{dk}, and then the observables of the theory, that are always dimensionless (as assumed in \cite{bdk,dk} and more recently remarked in the introductory part of references \cite{am,bm}), are not affected by this transformation that affects only dimensional quantities. In this sense the comparison between the members inside of a class is a non well-posed comparison \footnotemark \footnotetext{A well-posed comparison is one that can be resolved by physical experimentation.} so, the discussion that takes place in the current literature about which frame is the physical one, may, in most cases, be meaningless. Only theories that belong to different classes can be consistently compared.

Another important aspect of the BD gravity we shall refer in this letter is connected with the conformal symmetry. As it has been shown in Ref. \cite{far}, the graviational part of the JF BD Lagrangian density $L^I[g,\phi]$ (eq.(1) without ordinary matter) is invariant in form under the conformal rescaling of the spacetime metric:

$$
\gab \rightarrow \tilde \gab=\phi^{2\alpha}\gab
\eqno{(6)}
$$

the field redefinition:

$$
\tilde \phi=\phi^{1-2\alpha}
\eqno{(7)}
$$

and the coupling constant redefinition:

$$
\tilde \omega=\frac{\omega-6\alpha(\alpha-1)}{(1-2\alpha)^2}
\eqno{(8)}
$$

with $\alpha \neq \frac{1}{2}$. However, when one considers the full Lagrangian density (eq.(1)) the conformal invariance is broken, unless the matter stress-energy tensor is symmetric ($\Tab=T_{ba}$) and it has a vanishing trace: $T \equiv T^n_n = 0$. In this last case the conservation equation: $\nabla^n T_{na} = 0$ (which contains the dynamics of matter) is conformally invariant.

For his part, when one considers the JF of {\it Class II BD} gravity; i.e., the JF GR given by (4), the full theory is invariant in respect to the transformations (6-8), so the conformal symmetry is preserved even in the presence of ordinary matter. In this case, the dynamical equation of matter: $\nabla^n T_{na} = \frac{1}{2} \phi^{-1}\nabla_a \phi T$ is invariant in form respect to (6-8). In particular the free-motion equation for an uncharged, spinless mass point given by:

$$
\frac{d^2x^a}{ds^2} = - \Gamma^a_{nm} \frac{dx^n}{ds} \frac{dx^m}{ds} - \frac{1}{2} \phi^{-1} \nabla_n \phi (\frac{dx^n}{ds} \frac{dx^a}{ds} - g^{an})
\eqno{(9)}
$$

is preserved under (6-8). This amounts saying that, unlike the JF of {\it Class I BD} gravity, the presence of ordinary matter in the JF of {\it Class II BD} gravity does not break the scale-invariance of the theory. Full derivation of the conformal invariance of (4) will be given elsewhere.

Summing up: BD gravity theories can be grouped into two classes attending the behaviour of the dimensionless gravitational coupling constant in these theories. JF and EF formulations in each class are experimentally indistinguishable; only theories in different classes can be consistently compared. When comparing the JF formulation of {\it Class I BD} gravity (just BD theory)  with the JF representation of {\it Class II BD} theory (JF GR), one finds that in the second case the presence of ordinary matter does not break the conformal invariance of the theory.

We thank David I. Santiago for pointing out a crucial error about a consideration on WEP violation in the original version of this letter, and MES of Cuba for financing.


\begin{thebibliography}{99}
\bibitem{sok} L. M. Sokolowski in 'Proceedings of the 14th International Conference on General Relativity and Gravitation', Florenze (Italy)1995, M. Francaviglia, G. Longhi, L. Lusanna, E. Sorace (eds.), 337(World Scientific, 1997), gr-qc/9511073.
\bibitem{mag} G. Magnano and L. M. Sokolowski, Phys. Rev. D \textbf{50}, 5039(1994).
\bibitem{fgn} V. Faraoni, E. Gunzig, P. Nardone, IUCAA 24/98, e-print gr-qc/9811047 (to appear in 'Fundamentals of Cosmic Physics').
\bibitem{mae} K. Maeda, Phys. Lett. B \textbf{186}, 33(1987).
\bibitem{gp} D. J. Gross, M. J. Perry, Nucl. Phys. B \textbf{226}, 29(1983).
\bibitem{ss} D. I. Santiago, A. S. Silbergleit, gr-qc/9904003 ( Submitted to Phys. Rev. D 15).
\bibitem{bdk} C. Brans and R. H. Dicke, Phys. Rev. \textbf{124}, 925(1961).
\bibitem{dk} R. H. Dicke, Phys. Rev.\textbf{125}, 2163(1962).
\bibitem{am} A. Albrecht, J. Mag\"{u}eijo, e-print astro-ph/9811018.
\bibitem{bm} J. D. Barrow, J. Mag\"{u}eijo, e-print astro-ph/9811072.
\bibitem{far} V. Faraoni, Phys. Lett. A \textbf{245}, 26(1998).
\end{thebibliography}
\end{document}